\documentclass[preprint]{vgtc}               

\ifpdf
  \pdfoutput=1\relax                   
  \usepackage{graphicx}                
  \DeclareGraphicsExtensions{.pdf,.png,.jpg,.jpeg} 
\else
  \usepackage{graphicx}                
  \DeclareGraphicsExtensions{.eps}     
\fi%

\graphicspath{{figures/}{pictures/}{images/}{./}} 

\usepackage{microtype}                 
\PassOptionsToPackage{warn}{textcomp}  
\usepackage{textcomp}                  
\usepackage{mathptmx}                  
\usepackage{times}                     
\usepackage{cite}                      
\usepackage{tabu}                      
\usepackage{booktabs}                  

\usepackage{tabto}
\usepackage{xspace}
\usepackage[official]{eurosym}

\newcommand{\sys}{\textit{Sabrina}\xspace}

\vgtccategory{Research}

\vgtcinsertpkg

\preprinttext{\textcopyright 2019 IEEE. DOI: 10.1109/VISUAL.2019.8933598}

\title{\sys: Modeling and Visualization of Financial Data over Time\\with Incremental Domain Knowledge}

\author{
        Alessio Arleo\thanks{
            \{alessio.arleo$\mid$roger.leite$\mid$silvia.miksch\}@tuwien.ac.at}\\ %
        \scriptsize TU Wien %
        \and Christos Tsigkanos\\ %
             \parbox{1.4in}{\scriptsize \centering TU Wien}
        \and Chao Jia\thanks{\{chao.jia$\mid$manfred.klaffenboeck$\mid$
        michael.wimmer\}@cg.tuwien.ac.at}\\ %
            \scriptsize TU Wien
        \and Roger A. Leite\\ %
             \scriptsize TU Wien     
        \and Ilir Murturi\thanks{\{schahram.dustdar$\mid$ilir.murturi$\mid$christos.tsigkanos\}@dsg.tuwien.ac.at}\\ %
             \scriptsize TU Wien %
        \and Manfred Klaffenb\"ock\\ %
            \scriptsize TU Wien             
        \and \hspace{-0.5cm}Schahram Dustdar\\ %
            \scriptsize TU Wien     
                \and Michael Wimmer\\ %
            \scriptsize TU Wien        
        \and \hspace{-1.1cm}Silvia Miksch\\ %
             \parbox{1.4in}{\scriptsize \centering \hspace{-1.1cm}TU Wien, IIASA}
        \and \hspace{-1.1cm} Johannes Sorger\thanks{sorger@csh.ac.at}\\ %
            \hspace{-1.1cm}\parbox{1.4in}{\scriptsize \centering Complexity Science Hub Vienna, IIASA}   
}

\abstract{
\vspace{-3px}
    Investment planning requires knowledge of the financial landscape on a large scale, both in terms of geo-spatial and industry sector distribution. There is plenty of data available, but it is scattered across heterogeneous sources (newspapers, open data, etc.), which makes it difficult for financial analysts to understand the big picture. In this paper, we present \sys, a financial data analysis and visualization approach that incorporates a pipeline for the generation of firm-to-firm financial transaction networks. The pipeline is
    capable of fusing the ground truth on individual firms in a region with (incremental) domain knowledge on general macroscopic aspects of the economy. \sys unites these heterogeneous data sources within  a uniform visual interface that enables the visual analysis process. In a user study with three domain experts, we illustrate the usefulness of \sys, which eases their analysis process.
} 

\CCScatlist{
  \CCScatTwelve{Human-centered computing}{Visu\-al\-iza\-tion}{Visualization application domains}{Visual Analytics};
  \CCScatTwelve{Computing methodologies}{Modeling methodologies}{}{}
}

\teaser{
  \centering
  \includegraphics[width=0.88\textwidth]{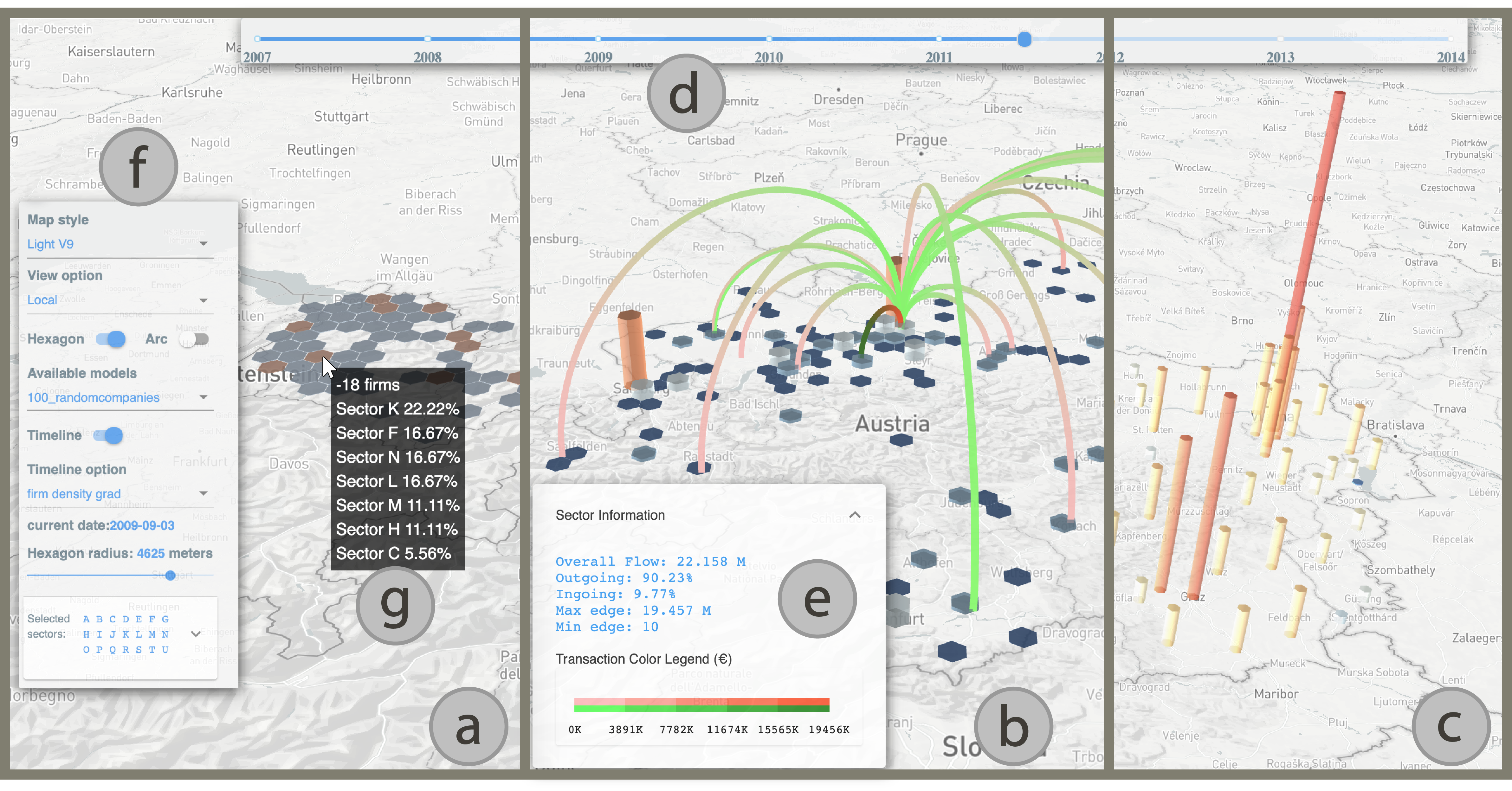}
    \vspace{-7px}
  \caption{The different visualizations composing the \sys system. The three views (a, b, c) display the firm density and transaction data of an Austrian economy dataset~\cite{sabinadataset}: (a) gradient encoding in blue/red indicates whether the firm density in a hexagonal area increased/decreased in respect to the previous time step; (b) inferred transactions between individual companies; (c) regionally clustered firm data. The interface allows the user to (d) navigate the temporal data dimension, (e) show aggregated details on selected transaction flows, and (f) adjust encoding and filtering options. Tooltips (g) show details on firms within a bin.
}
	\label{fig:teaser}
}

\begin{document}



\maketitle

\section{Introduction}
\begin{figure}[t!]
        \centering
        \includegraphics[width=0.95\linewidth]{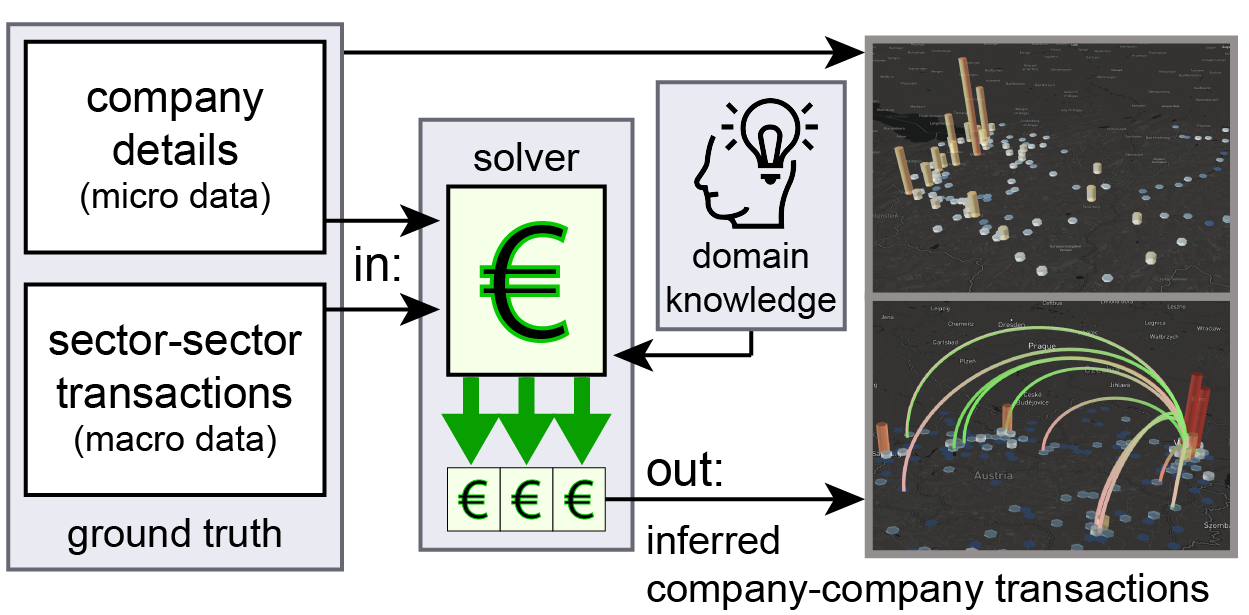}
        
        \caption{Sabrina data pipeline. The ground truth consisting of micro and macro financial data can be displayed on the map on different aggregation levels. Company-to-company transactions are inferred from the ground truth and domain knowledge that represent three types of input for an SMT solver.}
        \label{fig:sabrina_flow}
      
    \end{figure}

Economists and financial analysts try to understand the workings of the financial system, searching for profitable investment strategies or policies that have a stabilizing effect on markets.
However, it is hard to gain an overview of large multivariate financial data. There is no lack of information about the state of the economy in newspapers, on the internet or as open-data: however, they often tell only part of the story. More granular information about companies (or firms) is available in specific proprietary datasets cultivated by profit oriented research institutions, but the \textit{transactions}, i.e., monetary flows between individual firms, are generally completely unavailable. Such detailed monetary flows represent an essential piece of information when it comes to understanding, e.g., in which regions it is better to buy or sell specific goods or services, or which sectors (e.g., agriculture, manufacturing) are most and least proficient and where. 
In this paper, we present \sys, a Visual Analytics (VA) approach that supports economists and financial analysts in their investigation of national economies over time. To support these target users, \sys unites several financial data sources within a unified visual interface.
Particularly, \sys yields firm-to-firm transaction network models, in the form of weighted directed graphs.
Such a model condenses different financial data sources: granular information on specific firms (i.e., location, sector, financial performance), high level monetary flows between industry sectors, and domain knowledge (e.g., financial reports of the overall investment to and from a region, investments in higher education and high-tech companies, ``hunches'').
Domain knowledge can be infused incrementally into the basic model, to include further angles or updated information. Temporal information (e.g., the change in GDP of a region from one year to the next) can be used to evolve the model (see Fig. \ref{fig:sabrina_flow}).

\sys thus acts as a visual interface to two types of data: first, spatio-temporal financial data that, on the lowest level, contains information on individual firms; and second, transaction networks between individual firms. 
\sys allows users to investigate the financial data on several levels of aggregation, i.e., from regions and sectors up to precise locations and individual firms, and across time.
Our contributions are as follows:
\begin{itemize}
    \item A design study in the field of economic data analysis;\vspace{-4px}
    \item \sys, a prototype VA approach for financial data exploration, resulting from our task analysis;\vspace{-4px}
    \item A pipeline for inferring financial transaction networks;\vspace{-4px}
    \item A user study of \sys with three experts in economic analysis and policy evaluation.
\end{itemize}


\section{Related Work}\label{se:rel}

We divide related work into two main parts: VA approaches and model construction with formal methods.
In the former, we focus on approaches related to network analysis and dynamic exchange behavior; in the latter
we discuss
works using satisfiability 
in financial application domains (i.e., financial markets and banks).
    
    \textbf{VA approaches}. 
   Kirland's work~\cite{1999_kirkland} is among the first 
   to investigate the benefits and different aspects of VA. In this work, 2D and 3D techniques are used to combine different discovery features for market data, using diverse visualization and interaction types to perform network monitoring, alarms detection, and pattern-detection tasks.
    In order to prevent bitcoin owners from loss, anomaly detection is proposed by Pham~et~al.~\cite{2016_pham}. This approach uses learning methods, like Unsupervised Support Vector Machines to support different types of graphs~(line plots and scatter plots) on its information placements.
    Dynamic exchanging networks analysis is routine for banks. VIS4AUI~\cite{2012_didimo} focuses on financial crime analysis such as money laundering fraud. This system presents the combination of an interactive line plot and a node-link diagram designed for touchscreen interfaces. Clustering algorithms support the analysis, and the visual exploration of the networks is facilitated by abstraction and elaboration interaction techniques.
    A recent approach by Leite~et~al.~\cite{2018_leite} proposes ``EVA'' (Event detection with Visual Analytics), a VA approach to identify fraudulent events based on bank transactions logs. EVA combines well-known visualization techniques with a profile-based detection algorithm, using real-world data and experts to evaluate the approach. 
    
    \textbf{Model Construction with Formal Methods}.
    Financial systems are becoming even more complex and unpredictable with the ever-increasing amount of data. 
    The distribution of resources among several sources is referred to as Multiagent Resource Allocation (MARA)~\cite{chevaleyre2006issues}. MARA has been widely accepted and applies to various application domains such as stock trading~\cite{luo2002multi}, manufacturing systems~\cite{sousa2003fabricare}, etc. 
     Formal verification methods aim to assist different stakeholders (i.e., policy authors, banks, economists) to deal with the ever-growing complexity of their systems. Recently, some approaches have been introduced in financial systems using formal verification methods as the underlying reasoning mechanism to produce more accurate output. Passmore et al.~\cite{passmore2017formal} indicate the need for the use of formal verification methods in financial practice and regulation; 
     fostering the design, implementation, and regulation of critical algorithms that run modern financial markets. 
     For the sake of completeness, we also report approaches for the identification of economic models through simulation, such as multi-agent resource allocation~\cite{rothe2015economics} or agent-based modeling~\cite{poledna2017economic}.  
    
\section{Sabrina}\label{se:sabrina}
In this section, we discuss the design of \sys. First, we describe the methodology that drove the development of our idea.
Subsequently, we discuss the two main components of our approach: the incremental model building through domain knowledge and the design of the visual interface.

\subsection{Users, Data, and Tasks}\label{se:tasks}
To design \sys, we followed the nested approach by Munzner~\cite{munzner2009nested} and took advantage of the ``Design Triangle'' by Miksch and Aigner~\cite{miksch2014matter}. We determine (i) who will be the users of the system, (ii) what is the data model, and (iii) which tasks are to be achieved by our solution. To fully understand the problem domain, we conducted guided interviews with two researchers, experts in model building for financial data: they guided us through the process and highlighted some of the pitfalls they encountered during their work. Most importantly, they shared their ideas of the tasks that a system such as \sys should achieve. We used this knowledge to assemble our design triangle as follows:

    \textbf{Users.} \sys targets experts in financial data analysis and economics. They assess the development of national and international markets and analyze the financial performance of individual firms as well as their impact on their geographical and industrial environment, in a qualitative and quantitative manner. They also monitor the effect of policies on the financial system. 
    They are proficient in statistical data analysis but have limited experience with model building and visual analysis.
    
    \textbf{Data.} \sys is designed to handle heterogeneous information on companies within a specific region over time in the form of a dynamic directed weighted network. Its nodes are individual companies associated with several attributes (``micro'' data), such as the geographical location (coordinates/address of headquarters), name, sector (e.g., agriculture, industry -- encoded according to~\cite{onace2008}), financial performance (e.g., cash flow, personnel expenses). The links and weights of the network (the firm-to-firm transactions model) are initially unknown, since this data is not publicly available. Representative transactions can be inferred (see Sec.\ref{se:model}). The links are weighted according to the estimated amount of the financial transaction. Verifiable publicly available information on macroscopic aspects of the economy (``macro'' data) represents the ground truth for the inferred transactions. An example for macro data are sector-to-sector transactions (monetary in\&out flows between industry sectors, i.e., ``IO Tables''~\cite{iot2015} -- see Fig.\ref{fig:sabrina_flow}). Both the micro and the macro data can contain temporal aspects about the evolution of the economy over time. 
    
    

    
    \textbf{Tasks.} In the context of \sys, we focus on user tasks related to analyzing a national economy. With the help of the experts, we identified the following set of tasks.
    \textbf{T1}: \textit{Economy Status Overview}, monitoring multiple financial indicators on multiple hierarchical levels, i.e., national vs. regional, or firm vs. industry sector level.
    The user thereby scouts for interesting regions, sectors, or companies to investigate.
    \textbf{T2}: \textit{Monitoring changes over time}. Users can navigate through time points to investigate the change of financial indicators on different aggregation levels. An essential question is to understand if a policy change in a certain year is reflected in the growth (or the demise) of companies.
    \textbf{T3}: \textit{Investigation of relations between entities} (regions, sectors, companies). The user investigates how the distribution of transactions between firms, regions, and sectors evolves over time.
    \textbf{T4}: \textit{Model Comparison}. A model might only infer transactions for a specific region, area, or time interval. In this task, the user goes through multiple firm-to-firm transaction models. The objective is twofold: first, to experiment with different model constraints based on domain knowledge and to evaluate the resulting changes to create new hypotheses. Second, to focus on regional or sectors-specific models based on constraints that might invalidate a larger model.
    \textbf{T5}: \textit{Infusing domain knowledge}. In order to fine-tune a model, the user should be able to feed new domain knowledge in form of constraints into the model solver (see Sec. \ref{se:model}).
    

\subsection{Incremental Domain Knowledge}\label{se:model}

Models of financial behavior can vary greatly, since they are typically dependent on domain knowledge, make use of proprietary data or observations or are bound to particular analyses.
We are not concerned with economics aspects, but assume such domain knowledge is available as an input to the system. 
\sys supports any type of model data but is built on a methodology where the financial analyst can infer a model by incrementally introducing domain knowledge or through integrating models obtained from other sources.
To leverage knowledge of domain experts and users, we treat the problem of inferring a monetary outflow model as constraint satisfaction, where the constraints encode domain knowledge. A model that satisfies the constraints is a valid model. We note that this approach fits our general context particularly well: (i) domain knowledge encoded as constraints may be updated or expanded at will, (ii)  users may introduce their own knowledge for their analysis case, and (iii) the dynamic introduction of constraints renders our approach independent from particular financial modeling techniques, which may be integrated as well as further domain information. In the following, we describe the model inference process: it consists of three steps -- after general bounds are introduced, available high-level knowledge is captured, and experts' information is further introduced.

\textbf{Domain Bounds}. Absolute truths that are universally applicable are encoded as constraints to the problem, and used in the model validation, e.g., a company may not have negative expenses on employee costs.

\textbf{High-Level Behavior}. Macroeconomic monetary flows, e.g., between industry sectors, are generally known and publicly available~\cite{timmer2015illustrated, iot2015}, typically obtained from census or tax data. 
For example, it may be known that the outflow of companies from the manufacturing sector to the ones in agriculture amounts to 32 billion \EUR{} in 2014~\cite{iot2015}.

\textbf{Expert Knowledge}. Domain-specific insights are introduced as further constraints to the problem. This step leverages the fact that the intended users are experts in their domain, may possess further proprietary data or models, or use the system in the context of a what-if analysis. For example, an analyst within the financial department of a regional government may know that firms operating in a given municipality with less than 100k employee expenditure trade with a maximum of 10 firms in another region. 
Note that this step can be performed incrementally, on a case-by-case basis and individually by each user. In a similar fashion, the output of, e.g., a MARA process may be integrated as expert knowledge~\cite{chevaleyre2006issues}; explicit values of firm transactions are encoded as constants. 
The steps above correspond to component sets of different constraints. To build the target firm-to-firm transaction model, the specified constraints are encoded as first-order logical formulae within satisfiability modulo theories (SMT)~\cite{barrett2018satisfiability}, using
quantifiers (over finite sets) and integer linear arithmetic for constraint specification.
SMT solving is a highly computationally intensive operation. However, the complexity of constraints (viewed as logical formulae) is generally low, and the model inference operation is not required to be online.  
A valuation that satisfies the resulting formula is a valid model.
The output of the process yields a weighted graph, where nodes are firms and edges capture the value of a monetary outflow from one firm to another. In general, there can be many different graph variants satisfying the same constraints; the more constraints are specified, the better the model approximates reality. This is an inherent limitation of the approach -- minimal constraints can result in a broad range of satisfiable models. For a comprehensive and extensive analysis of the modeling process, the interested reader can refer to~\cite{iri}.

To evaluate the advocated model inference approach, we developed tool support and a proof-of-concept implementation based on the CVC4 SMT solver~\cite{cvc4}.
The resulting assignment is used to derive the firm transaction model. For reference, a problem instance involving 500 firms in Austria, taken from an anonymized subset of the \textit{Sabina} dataset~\cite{sabinadataset} 
is solved in 157 minutes on a laptop computer with an Intel i5 2.3GHz processor and 15G RAM. It is worth noting that how the constraints are expressed may have a significant impact on the running times~\cite{iri}.


\subsection{Visualization}

\textbf{System Overview.} \sys follows an overview + details-on-demand approach (\textbf{T1}). There are two main views, ``\textit{local}'' and ``\textit{regional}'' (see Fig.\ref{fig:teaser}). 
In the local view (Fig.\ref{fig:teaser}a,b), firms are spatially aggregated on an adjustable hexagonal grid. 
The binding between a hexagon and a firm is dynamic, therefore zooming-in (out) yields a more granular (coarse) visualization (i.e., there will be more/less hexagons that will contain fewer/more firms), while zooming out provides a coarser view. The height and the color of the hexagons encapsulate a user-defined aggregate value of one specific parameter regarding the contained firms. In this prototype, the available parameters are the number of the firms within a cluster and the aggregate cash flow (i.e., the sum of the money earned plus the money spent), both of which were deemed relevant by the researchers interviewed during the design of the prototype (see Sec. \ref{se:tasks}). 
To enable a time-dependent analysis, the user can switch the color encoding of hexagons to the gradient of the attribute change over time (\textbf{T2}). In this case, the color of the hexagons will change from blue to red depending on whether there was a decrease or increase in the selected value compared to the previous year (Fig.\ref{fig:teaser}a). Here, the height represents the amount of the change, allowing users to understand at a glance both the most (and least) proficient regions and the extent of the change. Time is navigated with the slider at the top of the map (Fig.\ref{fig:teaser}d); the change of the values of the current selected parameter between time steps (\textbf{T2}) is animated by linear interpolation.
By clicking on a hexagon, the in- and out-going transactions of the firms within the selected cluster are visualized 
as a ``Flow map''~\cite{jenny2018design} (\textbf{T3}), as displayed in Fig. \ref{fig:teaser}b. The color encodes both the direction of the transaction and its weight, i.e., the amount. The edge is colored in red (representing the ``loss'' of money) close to the transaction source and in green (representing the ``gain'') close to the target. 
The amount is encoded in the saturation of the color, following a continuous relative scale, making easy to understand how that specific edge compares with the others. 

When clicking, a details panel placed at the bottom left corner of the display is updated with statistics about the transactions (\% of funds going inwards/outwards, overall flow, transaction flow color legend) in the selected hexagon (Fig.\ref{fig:teaser}e). Hovering works both on hexagons and on edges, displaying context-sensitive tool-tips with aggregate value of firm density or cash flow and the contribution of each single sector in percent, and the weight of an edge respectively. When selected, a hexagon is kept selected while zooming in/out or when moving across time, with arcs and statistics updated accordingly. Since a firm-to-firm transaction model might not span across all the nodes in the current model, the user can decide to hide the firms that are not represented in the current one (i.e. firms that do not receive or issue any transaction in this model). This is particularly useful when an investigation has to be carried out on a specific subset of firms. Several transaction models can co-exist in the database, and it is possible to change between them (\textbf{T4}) through a menu on the configuration pane at the left side of the screen (Fig.\ref{fig:teaser}f).
The regional view follows the same design rationale as the local view, but the firm grouping is static and represents either the region, the city, the district, or other administrative aggregations depending on user selection (Fig.\ref{fig:teaser}c). Here, it is possible to switch between the visualization of absolute and area-normalized values (firms/km$^2$).
\sys is implemented as a web application written in JavaScript and leverages WebGL using the Deck.gl~\cite{keplergl} framework. A demo video is available at~\cite{sabrina}. 


\textbf{Design Rationale.} On an abstract level, the data on which \sys is operating consists of nodes (i.e., individual companies) and edges (i.e., the transaction links between them).
Our interviewed experts deemed the geo-spatial context of the data as important for the analysis, since the analysts are often interested in regional characteristics and patterns. 
The resulting requirement to show both firms and the transactions between them in the context of a map resulted in two challenges, which drove our design rationale.
First, (\textbf{C1}) the placement of firm data (nodes) according to their geographical coordinates results in severe over-plotting and thus clutter in densely populated areas.
Second, (\textbf{C2}) the rendering of transaction network edges between geo-spatial coordinates on top of the map results in the occlusion of the underlying nodes.
To address C1, we decided to apply hexagonal clustering (or ``binning'') on geo-spatial firm locations -- as a common technique in Geographical Information Systems (GIS) for aggregating individual point locations into polygonal regions. Hexagons 
achieve a more accurate approximation of large geo-spatial areas, display neighbourhood between cells clearly, and they are well suited when aspects of connectivity and/or movement paths have to be considered in the analysis~\cite{birch2007rectangular}. 
To address C2, we decided to render the map in a 2.5D context, enabling a pitched view where hex-bins are rendered on the x/z plane. Transaction links are drawn as a ``Flow map''~\cite{jenny2018design}, i.e., arcs that are bent towards the y-axis -- resulting in reduced occlusion between bins of companies and the transaction edges between them.
Additionally, the use of the vertical dimension intends to reduce edge crossings~\cite{vrotsou2017interactive}. Flow maps have been proven to be more effective than straight links in conveying flow direction and volume~\cite{dong2018using}, and precedent work investigated their use on a geographical context~\cite{rae2011flow}.
The trade-off of a 2.5D view are perspective distortion and occlusion, however the view angle can be dynamically adjusted to alleviate both issues.





\section{Evaluation: User Study}\label{se:eval}

In the scope of this short paper, we chose a qualitative evaluation as the validation method.
We invited three experts from the Austrian Chamber of Commerce for a two-hour discussion. The experts were enrolled in the validation because of their interest in \sys and their compatibility with the scope of the project. 
We followed a written protocol describing the sequence of questions and features to be presented; furthermore, we recorded the interview and the interactions on-screen during the entire demo session for post-evaluation transcription. Our goal was to understand whether the tasks that drove \sys's development were fulfilled;
moreover, we wanted to assess the validity of our assumptions and design choices.
The evaluation protocol consisted of three sequential parts. 

%
\textbf{Phase 1:} we interviewed the domain experts in order to learn about their workflow, i.e., their tasks, the data they investigate, and the tools they use to analyze them. As analysts of economic and trade policies, their main goal is the creation of better policy conditions for fostering entrepreneurship. Their main tasks are thereby the analysis of the data they gather from their members (+500K firms), and the presentation of gathered insights to peers. Examples for specific tasks are demographic analyses of companies (i.e., the members of the chamber), the evaluation of trade data, and the effect of policy changes on different levels (e.g., government level and firm level). For quantitative analyses, common data science software environments such as Python and R are used. Interestingly, their decision to participate in our evaluation was also based on the need for a solution to investigate and cross-reference data on their members in order to gain insights into the market, based on which decisions and policy changes can be inferred. These tasks overlap with \sys's (see Section \ref{se:tasks}), which is first partial evidence that the tasks identified during the initial design stage fit the target users. 

\textbf{Phase 2:} we introduced our system. We started by explaining the goal of \sys, followed by a scripted live demo, where we presented its features. Afterwards, we started a free exploration session of the system and the underlying data, where we guided the experts in issuing queries and investigating the inferred models.

\textbf{Phase 3:} we asked the experts about their feedback on \sys. The experts told us that previously, they did not have tools to visually investigate their data, and praised our idea of providing an overview of regional firm distributions and the associated financial values. The experts also appreciated the option to filter data by sector and to aggregate information by regions and observe their evolution over time (\textbf{T1}, \textbf{T2}). The interface (with some exceptions described below) was deemed immediate and effective. The infusion of additional domain knowledge into \sys's underlying transaction inference model was regarded as very interesting: the promise of investigating potential business relations sparked a great deal of interest in our experts (\textbf{T5}, \textbf{T3}). For instance, our model would allow the experts to differentiate between regions of high incoming and outgoing monetary flows for each sector and between different sectors. Moreover, they praised the feature of quickly switching from one model to the other to focus on firms belonging to a specific region (\textbf{T4}).



The experts also pointed out some limitations of \sys.
In the current version of \sys, we focused on displaying information on company headquarters; however, also the branch distribution of companies would be of interest to the experts. Finally, the experts expressed interest in a way to visually modify the constraints that generate a transaction model -- a feature that is not yet supported by our prototype, as the model specifications are currently scripted outside of the visual interface.
The overall feedback for \sys was very encouraging and indicates that our system generally fulfills the tasks presented in Sec. \ref{se:tasks}.

\section{Conclusions and Future Work}\label{se:conclusion}

In this paper we presented \sys, an exploratory work to create a system for the visualization of micro and macro economic data coupled with a pipeline to infer firm-to-firm transaction networks using incremental domain knowledge. A validation through expert interviews backed our design choices. 
Fueled by positive feedback, future research efforts will build on this foundation by targeting the development of tools for online model modification 
(\textbf{T5}), an improved design for model comparison (\textbf{T4}), means that enable cross-referencing of regional census data with firm data (e.g., ressources and infrastructures), and additional linked numerical views to complement the current sector information.




\acknowledgments{This work was partially supported by the Research Cluster ``Smart Communities and Technologies (Smart CT)'' at TU Wien, as well as the FFG Project 857136 at CSH Vienna.}

\bibliographystyle{abbrv-doi}

\bibliography{Sabrina}

\begin{thebibliography}{10}

\bibitem{keplergl}
kepler.gl.
\newblock \url{https://github.com/keplergl/kepler.gl}.
\newblock Accessed: 2019-06-12.

\bibitem{sabrina}
Sabrina: Modeling and visualization of financial data over time with
  incremental domain knowledge | centre for visual analytics science and
  technology.
\newblock \url{https://www.cvast.tuwien.ac.at/smartct/sabrina}.
\newblock Accessed: 2019-07-25.

\bibitem{sabinadataset}
Wirtschaftsuniversität wien: Sabina - info - datenbanken.
\newblock
  \url{https://www.wu.ac.at/bibliothek/recherche/datenbanken/info/sabina/}.
\newblock Accessed: 2019-06-10.

\bibitem{onace2008}
Önace 2008 - struktur.
\newblock \url{http://www.statistik.at/KDBWeb/kdb_VersionAuswahl.do}.
\newblock Accessed: 2019-07-20.

\bibitem{cvc4}
C.~Barrett, C.~L. Conway, M.~Deters, L.~Hadarean, D.~Jovanovi{'{c}}, T.~King,
  A.~Reynolds, and C.~Tinelli.
\newblock {CVC4}.
\newblock In G.~Gopalakrishnan and S.~Qadeer, eds., {\em Proceedings of the
  23rd International Conference on Computer Aided Verification (CAV '11)}, vol.
  6806 of {\em Lecture Notes in Computer Science}, pp. 171--177. Springer, July
  2011.
\newblock Snowbird, Utah.

\bibitem{barrett2018satisfiability}
C.~Barrett and C.~Tinelli.
\newblock Satisfiability modulo theories.
\newblock In {\em Handbook of Model Checking}, pp. 305--343. Springer, 2018.

\bibitem{birch2007rectangular}
C.~P. Birch, S.~P. Oom, and J.~A. Beecham.
\newblock Rectangular and hexagonal grids used for observation, experiment and
  simulation in ecology.
\newblock {\em Ecological modelling}, 206(3-4):347--359, 2007.

\bibitem{chevaleyre2006issues}
Y.~Chevaleyre, P.~E. Dunne, U.~Endriss, J.~Lang, M.~Lemaitre, N.~Maudet,
  J.~Padget, S.~Phelps, J.~A. Rodriguez-Aguilar, and P.~Sousa.
\newblock Issues in multiagent resource allocation.
\newblock {\em Informatica}, 30(1), 2006.

\bibitem{2012_didimo}
W.~Didimo, G.~Liotta, and F.~Montecchiani.
\newblock Vis4aui: Visual analysis of banking activity networks.
\newblock In P.~Richard, M.~Kraus, R.~S. Laramee, and J.~Braz, eds., {\em
  GRAPP/IVAPP}, pp. 799--802. SciTePress, 2012.

\bibitem{dong2018using}
W.~Dong, S.~Wang, Y.~Chen, and L.~Meng.
\newblock Using eye tracking to evaluate the usability of flow maps.
\newblock {\em ISPRS International Journal of Geo-Information}, 7(7):281, 2018.

\bibitem{jenny2018design}
B.~Jenny, D.~M. Stephen, I.~Muehlenhaus, B.~E. Marston, R.~Sharma, E.~Zhang,
  and H.~Jenny.
\newblock Design principles for origin-destination flow maps.
\newblock {\em Cartography and Geographic Information Science}, 45(1):62--75,
  2018.

\bibitem{1999_kirkland}
J.~D. Kirkland, T.~E. Senator, J.~J. Hayden, T.~Dybala, H.~G. Goldberg, and
  P.~Shyr.
\newblock The nasd regulation advanced-detection system (ads).
\newblock {\em AI Magazine}, 20(1):55, 1999.

\bibitem{2018_leite}
R.~A. Leite, T.~Gschwandtner, S.~Miksch, S.~Kriglstein, M.~Pohl, E.~Gstrein,
  and J.~Kuntner.
\newblock Eva: Visual analytics to identify fraudulent events.
\newblock {\em IEEE transactions on visualization and computer graphics},
  24(1):330--339, 2018.

\bibitem{luo2002multi}
Y.~Luo, K.~Liu, and D.~N. Davis.
\newblock A multi-agent decision support system for stock trading.
\newblock {\em IEEE network}, 16(1):20--27, 2002.

\bibitem{miksch2014matter}
S.~Miksch and W.~Aigner.
\newblock A matter of time: Applying a data--users--tasks design triangle to
  visual analytics of time-oriented data.
\newblock {\em Computers \& Graphics}, 38:286--290, 2014.

\bibitem{munzner2009nested}
T.~Munzner.
\newblock A nested model for visualization design and validation.
\newblock {\em IEEE transactions on visualization and computer graphics},
  15(6):921--928, 2009.

\bibitem{passmore2017formal}
G.~O. Passmore and D.~Ignatovich.
\newblock Formal verification of financial algorithms.
\newblock In {\em International Conference on Automated Deduction}, pp. 26--41.
  Springer, 2017.

\bibitem{2016_pham}
T.~Pham and S.~Lee.
\newblock Anomaly detection in bitcoin network using unsupervised learning
  methods.
\newblock {\em arXiv preprint arXiv:1611.03941}, 2016.

\bibitem{poledna2017economic}
S.~Poledna, M.~Miess, and S.~Thurner.
\newblock Economic forecasting with an agent-based model.
\newblock 2017.

\bibitem{rae2011flow}
A.~Rae.
\newblock Flow-data analysis with geographical information systems: a visual
  approach.
\newblock {\em Environment and Planning B: Planning and Design},
  38(5):776--794, 2011.

\bibitem{rothe2015economics}
J.~Rothe.
\newblock {\em Economics and computation}, vol.~4.
\newblock Springer, 2015.

\bibitem{sousa2003fabricare}
P.~Sousa, C.~Ramos, and J.~Neves.
\newblock The fabricare scheduling prototype suite: Agent interaction and
  knowledge base.
\newblock {\em Journal of Intelligent Manufacturing}, 14(5):441--455, 2003.

\bibitem{timmer2015illustrated}
M.~P. Timmer, E.~Dietzenbacher, B.~Los, R.~Stehrer, and G.~J. De~Vries.
\newblock An illustrated user guide to the world input--output database: the
  case of global automotive production.
\newblock {\em Review of International Economics}, 23(3):575--605, 2015.

\bibitem{iot2015}
M.~P. Timmer, E.~Dietzenbacher, B.~Los, R.~Stehrer, and G.~J. de~Vries.
\newblock An illustrated user guide to the world input–output database: the
  case of global automotive production.
\newblock {\em Review of International Economics}, 23(3):575--605, 2015. doi:
  {{%
10\hspace{.1pt}\discretionary{.}{%
}{.}\hspace{.4pt}1111\discretionary{/}{%
}{/}roie\hspace{.1pt}\discretionary{.}{%
}{.}\hspace{.4pt}12178}}


\bibitem{iri}
C.~Tsigkanos, A.~Arleo, J.~Sorger, and S.~Dustdar.
\newblock How do firms transact? {{Guesstimation}} and validation of financial
  transaction networks with satisfiability.
\newblock In {\em {IEEE} 20th International Conference on Information Reuse and
  Integration, Los Angeles, California, USA, July 30 - August 1 2019, (to
  appear)}.

\bibitem{vrotsou2017interactive}
K.~Vrotsou, G.~Fuchs, N.~Andrienko, and G.~Andrienko.
\newblock An interactive approach for exploration of flows through
  direction-based filtering.
\newblock {\em Journal of Geovisualization and Spatial Analysis}, 1(1-2):1,
  2017.

\end{thebibliography}
\end{document}